\documentclass[acmsmall, manuscript]{acmart}
\AtBeginDocument{%
  }

\setcopyright{acmlicensed}
\copyrightyear{2018}
\acmYear{2018}
\acmDOI{XXXXXXX.XXXXXXX}

\acmJournal{JACM}
\acmVolume{37}
\acmNumber{4}
\acmArticle{111}
\acmMonth{8}

\begin{document}

\title{RefactorAssist: Agentic Refinement for Reliable Code Refactoring}

\author{Jonathan Cordeiro}
\email{19jac16@queensu.ca}
\orcid{0009-0005-9954-3359}
\affiliation{%
  \institution{Queen's University}
  \city{Kingston}
  \country{Canada}
}

\author{Shayan Noei}
\email{s.noei@queensu.ca}
\orcid{0000-0002-5675-7817}
\affiliation{%
  \institution{Queen's University}
  \city{Kingston}
  \country{Canada}}
\email{}

\author{Ying Zou}
\email{ying.zou@queensu.ca}
\orcid{0000-0002-5335-0261}
\affiliation{%
  \institution{Queen's University}
  \city{Kingston}
  \country{Canada}
}

\renewcommand{\shortauthors}{Cordeiro et al.}

\begin{abstract}
Code refactoring aims to enhance the internal structure of source code without affecting its functional behavior. 
The recent advancements of Large Language Models (LLMs) have demonstrated potential for automating software engineering tasks, such as code refactoring, offering new opportunities to improve code quality automatically.
However, the refactorings produced by LLMs often introduce subtle errors, leading to functional behavior changes and failed unit tests, which limit their practical adoption. 
To address the limitations of LLM-generated refactorings, we analyze the root causes of their failures and develop the RefactorAssist agent to improve the functional correctness of LLM-generated refactorings. To this end, we use 10 open-source Java projects with their native test suites, and we first manually evaluate why LLM-generated refactorings fail unit tests. We then design an automated agentic approach that leverages unit-test logs, error explanations, project context retrieval, and code diffs to guide the iterative refactoring process.
Our findings show that the main reasons for refactorings failing unit tests are context misunderstanding/hallucination (24.3\%), incorrect or inconsistent renaming (15.3\%),  adding new functionality or variables (13.7\%), code incompleteness (11.3\%), syntax and structural errors (9.7\%), edge cases not handled (9\%), improper type handling (8.7\%), and variables outside scope (8\%). To make our approach cost-effective, our RefactorAssist agent first applies a static repair step that corrects syntactic issues of the code, such as missing imports, unbalanced brackets, and compilation errors, before test execution without LLMs. For refactorings that still fail test cases, RefactorAssist incorporates error logs and code diffs, achieving up to a 70.8\% repair rate on the remaining failures and up to a 94.2\% cumulative pass rate under the best-performing configuration. These results indicate that static checks as well as test-guided, context-aware agentic repair can increase the reliability of LLM-generated refactorings, bringing them closer to practical integration within developer workflows.
\end{abstract}




\maketitle

\section{Introduction}
\label{sec:introduction}
Code refactoring is the process of restructuring existing code to enhance its quality while preserving its external functionality~\cite{fowler2018refactoring, fowler1999refactoring}. Effective refactoring not only improves readability and modularity but also reduces dependencies and addresses technical debt~\cite{Refactoring.Guru, noei2023empirical}. Although human developers perform better in code refactoring due to their contextual awareness and domain expertise~\cite{fowler1999refactoring}, they also pose the risk of human errors~\cite{anu2018development, cordeiro2025llm}. Therefore, automated approaches can reduce the dependence on developers and save time. Classical static tooling, including general code smell and bug detectors such as SonarQube~\cite{sonarqube_software}, PMD~\cite{pmd_software}, and SpotBugs~\cite{spotbugs_software}, and code smell detection tools such as DesigniteJava~\cite{sharma2016designite}, can flag design problems and suggest limited refactoring types. However, their adoption at scale is limited by noisy alerts and false positives, time-intensive configuration, and no guarantee that suggested changes preserve functional behavior. Studies report under-use or suppression of alerts in practice, sometimes accruing technical debt~\cite{murphy2008refactoring, johnson2013don, lenarduzzi2023critical}. 

The advancement of automatic code generation has revolutionized software development, enabling developers to produce code more efficiently and consistently~\cite{svyatkovskiy2020intellicode, fan2023large}. Despite the capabilities of Large Language Models (LLMs) in automating a variety of programming tasks, including code generation and bug fixing~\cite{wei2023copiloting, fan2023large}, code changes related to software design and architecture, such as refactoring, remain unreliable~\cite{liu2024empiricalstudypotentialllms}. LLM-generated refactorings often introduce unintended behavioral changes, violate architectural constraints, or generate refactorings that fail existing test cases~\cite{cordeiro2025llm, cordeiro2024empirical, chang2024survey, Huang_2024}, highlighting the need for specialized refactoring agents designed specifically for safe and reliable refactorings. 

In this study, we first empirically evaluate how often LLMs fail to perform refactorings and investigate the underlying causes of these failures. We then propose and evaluate an agentic approach (i.e., RefactorAssist) for repairing failed LLM-refactored code. RefactorAssist first applies a static repair that fixes syntax and structural issues in LLM-refactored code to reduce the cost of running LLMs. It then employs an iterative agentic repair loop to resolve functional issues introduced by LLM-generated refactorings. RefactorAssist combines compilation feedback, unit-test logs, code diffs, retrieved project context, and diagnostic explanations to guide refactoring repair, and recover from failures introduced by LLM-generated refactorings.

To demonstrate the effectiveness of our approach, we utilize the Microsoft Methods2Test dataset~\cite{Tufano_2022}, a collection of real-world Java methods paired with their original unit tests extracted from mature open-source projects, to develop and test RefactorAssist. To empirically evaluate our design of RefactorAssist, we apply it to incorrect LLM-generated refactorings and measure whether it can repair the introduced functional failures while preserving the intended refactoring behavior. To minimize data leakage and better reflect real-world settings where refactoring tasks are unseen by the model, we use StarCoder2~\cite{lozhkov2024starcoder2stackv2} for refactoring generation. StarCoder2 is open-source and trained on a publicly available Stack v2 dataset~\cite{lozhkov2024starcoder2stackv2}, allowing us to verify that the evaluation repositories are not included in its training. 

In summary, our work addresses the following research questions:

\textbf{RQ1: How often does LLM-based refactoring lead to incorrect code behavior?}  
Although LLMs demonstrate strong capabilities in code generation and refactoring, their functional correctness remains a significant concern. A refactoring must preserve external behavior; therefore, transformations that improve code structure but introduce compilation errors or failing tests cannot be considered correct refactorings. To evaluate this issue, we measure the unit test pass rates of refactorings generated by state-of-the-art LLMs, including StarCoder2, GPT-4o, Claude Sonnet, and Qwen2.5-Coder-32B-Instruct, under 0-shot, 1-shot, 3-shot, and 5-shot prompting settings. We compile each refactored program and execute the associated unit tests to determine whether the generated refactoring preserves behavior. Our findings show that model choice has the strongest effect on functional correctness, with GPT-4o achieving the highest overall pass rate of 80.8\%. In contrast, few-shot prompting and refactoring scope have limited effects on correctness, indicating that simply adding examples or restricting the transformation scope is insufficient to ensure reliable refactoring generation.

\textbf{RQ2: What are the most common reasons for failures in LLM-refactored code?}
To understand the root causes of unit test failures, we first apply an LLM-assisted clustering approach to cluster all failure messages into distinct failure categories. We then conduct a manual analysis on a statistically representative sample of failures to uncover recurring patterns and verify the failure patterns. Both analyses converge on the same fixed taxonomy: (1) context misunderstanding/hallucination, (2) new functionality introduced, (3) incorrect or inconsistent renaming, (4) scope/dependency conflicts, (5) improper type handling, (6) code incompleteness, (7) unhandled edge cases, and (8) syntax/structural errors. Our analysis reveals that the causes of unit test failures are context misunderstanding/hallucination (24.3\%), incorrect or inconsistent renaming (15.3\%),  adding new functionality or variables (13.7\%), code incompleteness (11.3\%), syntax and structural errors (9.7\%), edge cases not handled (9\%), improper type handling (8.7\%), and variables outside scope (8\%).

\textbf{RQ3: Can our agentic approach improve the correctness of LLM-refactored code?}  
The failure analysis in the second research question indicates that many incorrect refactorings stem from syntactic and structural issues, while others require deeper semantic diagnosis and repair. To address both types of failures, we introduce RefactorAssist, which combines a repair pipeline in which static fixes are applied first, followed by an agentic refactoring repair stage for the remaining failing cases. The static-fix stage preserves imports, enforces bracket matching, and performs basic type-consistency checks before test execution, while the agentic approach uses compiler and unit-test failure logs, code diffs, and an error diagnostic LLM explanation to guide targeted repair generation. Our results show that this combined approach substantially improves the overall unit test pass rate from 68.4\% to 80.3\% after static intervention and further rises to 93.4\% after utilizing our agentic approach.

\textbf{RQ4: Which components of our approach contribute most to repair effectiveness?}  
To understand which components of RefactorAssist drive the most improvement in refactoring correction, we perform an ablation study that evaluates the contribution of different components of RefactorAssist. We find that the highest-performing configuration combines the code diff, error log, and LLM explanation output from the GPT-4o error diagnostic LLM. This configuration repairs 70.8\% of the refactorings, contributing to a cumulative unit test pass rate of 94.2\%. In contrast, adding retrieval-augmented context has a negative effect on the cumulative pass rate. 
The key contributions of this work are as follows:
\begin{itemize} 
    \item An evaluation of LLM-generated refactorings using the public Methods2Test dataset, measuring unit test failure rates to assess their reliability in preserving functional behavior.
    \item An analysis of common failure reasons in LLM-refactored code, providing insights into recurring issues along with quick static fixes.
    \item The development of an agentic approach to refine LLM refactorings, improving their functional correctness.
    \item We provide a replication package to enable the reproducibility of our study. The replication package of the study can be accessed at: \url{https://github.com/Software-Evolution-Analytics-Lab-SEAL/Automating-refactoring-correction}
\end{itemize}

\textbf{Paper Organization.} The remainder of this paper is structured as follows. Section~\ref{sec:method} describes RefactorAssist and its repair workflow. Section~\ref{sec:experiment_setup} outlines the research approach and methodology. Section~\ref{sec:results} provides the results of the research. Section~\ref{sec:threats_to_validity} discusses the threats to validity. Section~\ref{sec:related_work} summarizes existing work on automatic refactoring, testing, and automated refactoring repair. Finally, Section~\ref{sec:conclusion} concludes the paper by summarizing our contributions and outlining future research directions.
\section{RefactorAssist}
\label{sec:method}
\begin{figure*}
    \centering
    \includegraphics[width=\linewidth]{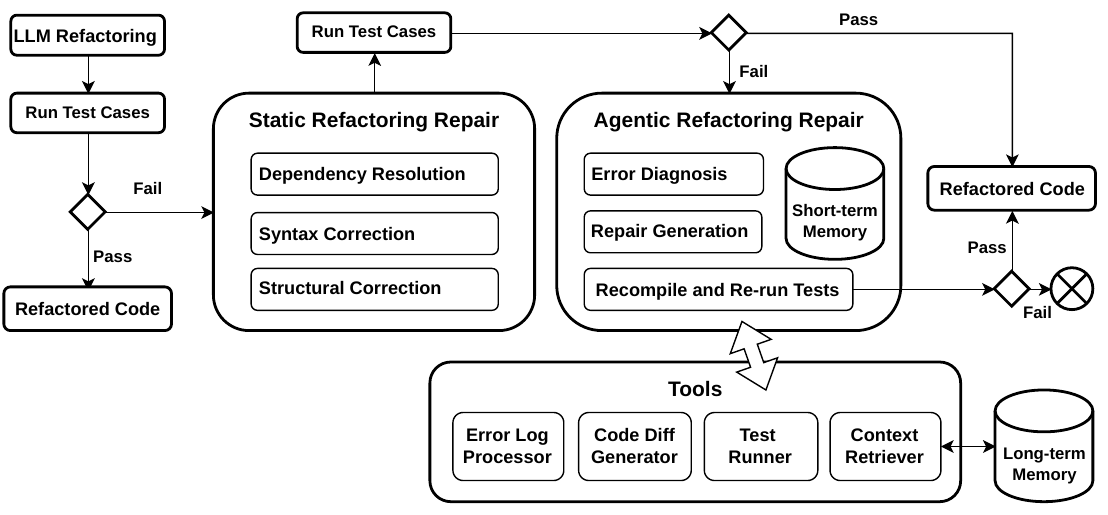}
    \caption{Overview of RefactorAssist. Failing LLM-generated refactorings first undergo static repair and, if still failing, enter an iterative agentic repair loop.}
    \label{fig:feedback_loop}
\end{figure*}

RefactorAssist is designed to improve incorrect LLM-generated refactorings. RefactorAssist first applies static fixes to address dependency, syntactic, and structural errors without additional LLM calls. If the refactoring still fails, it enters an iterative agentic repair loop that uses failure evidence, project context, and diagnostic explanations to guide targeted repair generation. Figure~\ref{fig:feedback_loop} provides an overview of RefactorAssist.

\subsection{Static Refactoring Repair}
The first repair phase of RefactorAssist applies static fixes before invoking the agentic repair loop. This phase is designed to address failure patterns that do not require deeper semantic reasoning.  The static repair phase aims to repair the initial LLM-generated refactoring that fails at least one of its associated unit tests. The static fixes are conservative and syntax-preserving and are intended to correct structural issues without altering the intended program logic. Specifically, the static repair stage performs the following operations:

\begin{itemize}
    \item \textit{Dependency Resolution:} This operation targets failures related to variables outside scope, dependency conflicts, and incomplete context, where the LLM removes imports that are still required by the refactored code. We extract the import declarations from the original file before refactoring and compare them with those in the LLM-refactored file. If an original import is removed during generation, we restore it in the refactored file as a conservative fix for omitted import statements.

    \item \textit{Syntax Correction:} This operation targets syntax and structural errors as well as incomplete code outputs, which are identified in the second research question as recurring causes of unit test failure. We apply a stack-based pass to ensure every \{, [, and ( has a corresponding closing counterpart. When the imbalance can be resolved through a single-token correction, the missing delimiter is inserted at the nearest safe location.

    \item \textit{Structural Correction:} This operation targets improper type handling and simple structural inconsistencies introduced during refactoring. Therefore, we check for mismatches, such as assigning a \texttt{String} to an \texttt{int} or producing incompatible method signatures, and apply patches when the correction is syntactically clear.
\end{itemize}

\subsection{Agentic Refactoring Repair}
The failed refactoring undergoes static refactoring repair and is re-evaluated using the associated unit tests. If the refactoring still fails, it enters the agentic repair loop of RefactorAssist, where it diagnoses the failure, generates a repair, recompiles the project, and reruns the tests until the refactoring passes or reaches the iteration limit. 

\subsubsection{Tools}
RefactorAssist uses four dedicated tools to collect failure evidence, retrieve project context, and construct repair guidance for the code generation LLM. These tools are described below:

\begin{itemize}
    \item \textbf{Error Log Processor:} This tool transforms raw compiler and unit-test output into structured failure evidence for downstream diagnosis and repair. To this end, for each failing refactoring, we capture the complete unit-test output, including compiler diagnostics, stack traces, and failed assertions. To keep the input consistent across projects, we standardize environment-specific details by replacing absolute file paths and temporary directory names with placeholders. We then extract two failure-relevant signals for downstream analysis: (1) an error signature, including the exception type, assertion message, and failing test name, and (2) an approximate failure location, including the class, method, and line ranges referenced in the logs.

    \item \textbf{Code Diff Generator:} We compute a unified diff between the original file and the LLM-refactored file to explicitly capture the code changes introduced by the refactoring. The diff is used to identify candidate patch regions by selecting diff hunks that overlap with the failure locations or reference symbols mentioned in the error logs, enabling analysis and repair of the modified code regions.

    \item \textbf{Context Retriever:} To provide the repair process with project context that is often missing from the failing file, RefactorAssist retrieves relevant dependency information from the same repository. The retrieval query is constructed from program symbols extracted from the error logs and code diff, including class names, method signatures, package imports, and referenced fields. These symbols are used to locate repository files and code snippets that define or constrain the failing code. The retrieved context includes dependent utility classes, type declarations, interfaces, constants, and method definitions that are referenced by the failing code or test. When the retrieved content exceeds the prompt-size limit, we prioritize the snippets most directly related to the symbols appearing in the error log and diff and remove unrelated methods, comments, or file regions. The goal of this tool is to expose the dependencies that the code generation LLM must preserve when proposing a fix.

   \item \textbf{Test Runner:} This tool compiles the modified project and executes the associated unit tests after each static-fix or repair attempt. The resulting pass/fail outcome determines whether the refactoring is accepted, passed to the agentic repair loop, or used as feedback for the next iteration.
\end{itemize}

\subsubsection{Memory}
RefactorAssist maintains both short-term and long-term memory. 
\textit{Short-term memory} is used at runtime and captures failure-specific evidence from the current iteration, including the latest unit-test error log and code diff. This information is updated after each repair attempt. \textit{Long-term memory} stores project-level context retrieved from the repository, such as relevant dependencies and type information, which is reused across repair iterations to provide repository-level context.

\subsection{Agent Runtime}
At the agentic refactoring repair runtime, RefactorAssist follows the two-stage repair process shown in Figure~\ref{fig:feedback_loop} and described below.

\subsubsection{Error Diagnosis}
The error diagnosis phase converts raw failure evidence into a structured diagnosis that the code generation LLM can use for repair. RefactorAssist first uses the \textit{Error Log Processor} tool to extract compiler and test failure information from the raw logs, including error coordinates, diagnostic messages, failed assertions, and optional \texttt{symbol:}/\texttt{location:} lines. It then uses the \textit{Code Diff Generator} tool to summarize the changes between the original file and the LLM-refactored file as a unified diff. Next, the \textit{Context Retriever} tool retrieves relevant project context and common fix patterns from long-term memory based on the symbols and failure locations appearing in the error log and diff.

Using these inputs, RefactorAssist constructs an ``explain'' prompt for the error diagnostic component. The error diagnostic component uses an LLM to explain the failure and suggest a repair plan. Its output is a tuple \{\textit{root\_cause}, \textit{hint}\}, where the root cause summarizes why the refactoring failed, and the hint provides a concise fix suggestion containing specific symbols, such as class, method, or parameter names. This tuple is then passed to the repair prompt in the subsequent stage.

\subsubsection{Repair Generation}
The repair generation phase converts the error diagnosis into a revised Java source file. To this end, the failing code, the code diff, and the plan generated by the error diagnosis component are assembled into a repair prompt and passed to the code-generation LLM to produce a revised version of the source code.

The repair prompt instructs the code-generation LLM to return only runnable Java source code and constrains the model to avoid unnecessary class or top-level type changes, preserve public signatures unless required by the failure, and keep edits minimal and localized to the regions implicated by the diff or error log. The goal is to convert the diagnostic explanation into a repair attempt that restores compilability and unit-test correctness while preserving the intended refactoring. The full prompt templates used for diagnosis and repair are provided in our replication package.

\subsection{Recompile and Re-run Tests}
After each repair attempt, RefactorAssist replaces the failing file with the repaired version and uses the \textit{Test Runner} tool to recompile the project and rerun the associated unit tests. If the repair succeeds, the process stops. Otherwise, the \textit{Error Log Processor} tool captures the new failure evidence, short-term memory is updated with the latest log, diff, and diagnostic information, and the next repair iteration begins. This loop continues for up to 10 iterations or until the refactored code passes all associated tests. We use this iteration limit to balance repair opportunity and efficiency, allowing multiple rounds of diagnosis and correction while reducing computational cost and limiting the risk of drifting away from the original refactoring intent.

\section{Experiment Setup}
\label{sec:experiment_setup}

This section describes the experimental setup of our study, including project selection, data preparation, and analysis methods.

\begin{figure*}[t]
    \centering
    \includegraphics[width=\linewidth]{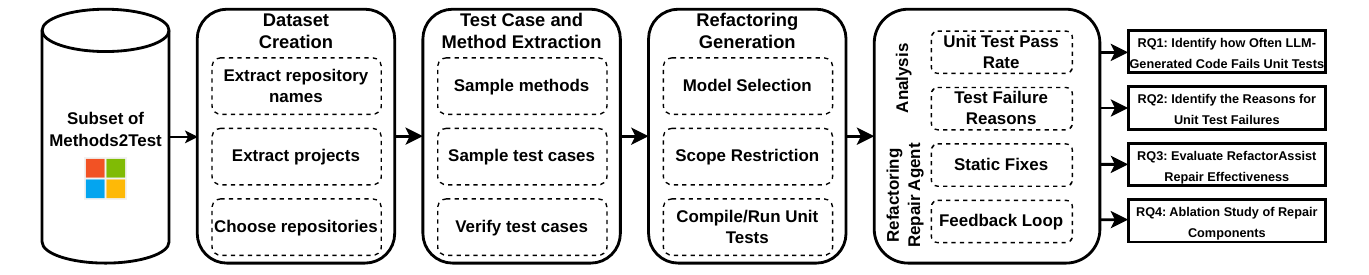}
    \caption{Overview of the research approach.}
    \label{fig:overview}
\end{figure*}

\subsection{Overview of Our Approach}
Figure~\ref{fig:overview} shows the overview of our study. We begin by collecting our dataset from the corresponding GitHub projects in the Methods2Test dataset~\cite{Tufano_2022}. For the initial refactoring generation, we utilize StarCoder2~\cite{lozhkov2024starcoder2stackv2} due to its open-source availability and transparent training data, which allows us to mitigate potential data leakage. Data leakage refers to the risk that the data used in our evaluation was already seen during model training, which can artificially inflate performance~\cite{Brownlee_2020, kapoor2023leakage}. Then each generated refactoring is compiled and executed against the unit tests of the original project to measure its functional correctness. When a refactored method fails the unit tests, we analyze the resulting test failures to understand common failure patterns and identify opportunities for improvement. Leveraging identified failure patterns, we implement a repair pipeline to rectify and refine the initial refactoring attempts. Therefore, we employ an agentic approach that first performs static fixes, then analyzes error logs and code diffs between the original and refactored code to identify and repair failures.

\subsection{Dataset Creation}
This section describes the construction of our evaluation dataset, including filtering steps to mitigate data leakage and ensure sufficient test coverage.

\subsubsection{Subject Selection}
\label{sec:project_selection}
We conduct our study using repositories from the Methods2Test dataset~\cite{Tufano_2022}, which contains 9,410 open-source Java repositories and metadata linking methods to their corresponding test cases. Since Methods2Test does not provide ground-truth refactorings required to compare against LLM-generated code, we use the repository names referenced by Methods2Test as our target projects. From these projects, we independently extract methods and their corresponding unit tests using our own parsing and mapping pipeline.

Our project selection approach is as follows:
\begin{enumerate}
    \item Data Leakage Prevention:
    To address potential data leakage, we extract all repository names from the Stack-v2 dataset, which forms the training corpus for StarCoder2~\cite{lozhkov2024starcoder2stackv2}. By cross-referencing these names against those in the Methods2Test dataset, we exclude all repositories that overlap. This filtering step reduces the dataset from 9,410 to 4,447 repositories and ensures that no evaluation repository was seen during training.
    
    \item Test Coverage Filtering:  
    From the remaining repositories, we select only those with at least 1,000 runnable unit tests~\cite{noei2025empirical}. This ensures that each selected project provides sufficient test coverage for reliable evaluation of behavioral correctness. After this filtering step, 17 repositories remain.
    
    \item Author and Redundancy Filtering:
    To increase project diversity and avoid potential author or organizational bias, we remove repositories developed by the same authors or organizations (e.g., forks, template variants). This step results in a final set of 10 unique repositories.
\end{enumerate}

Our subject selection process aims to balance experimental rigor and evaluation scalability while preserving project diversity, test coverage, and protection against potential data leakage. In our agentic approach, each refactoring undergoes up to 10 feedback iterations across 11 ablation configurations, resulting in thousands of compile and test cycles per project. Running these experiments demands substantial multi-GPU resources, as each iteration involves repeated model inference, compilation, and unit test execution. Even when utilizing a cluster with eight NVIDIA A100 GPUs, expanding beyond this scale would be computationally infeasible within a reasonable experimental time frame.

\subsubsection{Instance Selection}
From our selected subjects, we randomly sample 1,000 passing method-test pairs from each of the 10 repositories, which provides 95\% confidence with a 5\% margin of error~\cite{xu2024dataefficientevaluationlarge, shao2024balanceddatasamplinglanguage, noei2023empirical}, resulting in a final evaluation suite of 10,000 unique method-test pairs across 10 diverse projects. This configuration is optimized to maximize architectural diversity while maintaining computational feasibility relative to the intensive requirements of the refactoring pipeline.
 
Furthermore, we implement a Python script that processes each method–test pair in the Methods2Test dataset. For each pair, we check whether the mapped test class actually corresponds to the target method and whether the test can be executed successfully in the original project environment. To do this, the script builds the project and runs the associated test class referenced by the dataset.
\begin{itemize}
    \item If the repository is not already available in the local workspace, the script clones it recursively together with its submodules. If a local copy already exists, the script updates it to the required version.

    \item The script then checks whether a top-level \verb|pom.xml| file exists. In Maven-based Java projects, the \verb|pom.xml| file defines the project structure, dependencies, build plugins, modules, and test configuration. Therefore, we use this file as the entry point for building the project with Apache Maven~\cite{Miller2010ApacheM}. When the file is present, the script builds the entire project so that all required modules and dependencies are installed locally before executing the associated test class.
\end{itemize}

After setup, each test class is executed in isolation using Apache Maven, with a timeout of 600 seconds per invocation to filter out invalid or non-terminating tests. The script captures both standard output and standard errors and uses the Maven test result to determine whether the test passes. Only method–test pairs whose associated tests can be built and executed successfully are retained for our study.

\subsection{Test Case and Method Extraction}
\label{sec:test_method_extraction}
To evaluate the functional correctness of LLM-generated refactorings, we map the test cases to their corresponding methods within the selected repositories. We create a script to automate the identification and linking of test cases to their associated methods. The process to extract only valid, runnable test cases begins with cloning each of the 10 selected repositories locally. All Java files within each repository are parsed using the JavaParser library~\cite{javaparser}, which analyzes both production-code and test-code directories to extract methods and relevant structural metadata; for each production method, we record its fully qualified name, including the package name, class name, and method name, as well as its file path, method signature, and the enclosing class or interface. For each test case, we similarly record the test method name, file path, enclosing test class, package name, and any invocation relationships that link the test to a target production method.

The methods in the test code are identified based on the presence of the \texttt{@Test} annotation, their location in files under the \texttt{src/test/} directory, or naming conventions commonly used in test frameworks, such as \texttt{Test} or \texttt{test}~\cite{junit5}. Once a test method is located, we examine its body to find the method calls it invokes. For example, a test method such as \texttt{testAddItem()} in \texttt{CartTest} may contain a call to \texttt{cart.addItem(product)}. In this case, the invoked method is resolved to its fully qualified name, such as \texttt{com.shop.Cart.addItem(Product)}, which we identify as the target production method under test. More generally, our goal is to determine which specific production method a test is intended to validate, which is what we refer to as the target method. To identify the target method, each method call in the test is resolved to its fully qualified name (\textit{i.e.,} class, package, and method signature), allowing us to accurately map a test to the exact method under test. Our extracted data contains the repository name, the fully qualified method name, the fully qualified test name, and the file paths of both the method and the test, which are included in our replication package.

\subsection{Refactoring Generation}
\label{sec:refactoring_generation}
The refactoring generation stage produces the initial LLM-generated refactoring for each target method or class. 
For each target, the selected LLM generates one refactored candidate per prompt configuration. We then compile the modified project and run the associated unit tests to determine whether the generated refactoring preserves functional correctness. Unit tests are a widely adopted mechanism to validate functional correctness and detect regressions; therefore, we examine the unit test pass rates of the initial LLM-refactored code. A refactoring is considered correct if it compiles successfully and all associated unit tests pass. We report this outcome using the unit test pass rate, defined as follows: 

\begin{equation}
\label{equ:pass rate}
\text{Pass Rate} = \frac{\text{\# of refactorings that compile and pass all tests}}{\text{\# of total refactorings}}
\end{equation}

\noindent We evaluate four LLMs for initial refactoring generation:
\begin{itemize}
    \item StarCoder2~\cite{lozhkov2024starcoder2stackv2} is included as an open-source code model with publicly documented training data, which makes it suitable for controlled experimentation and data-leakage mitigation.

    \item GPT-4o~\cite{openai2024gpt4} is selected as a strong closed-source model to represent the performance of modern commercial LLMs on behavior-preserving refactoring.

    \item Claude Sonnet~\cite{Anthropic2024Claude3.5Sonnet} is included as another closed-source commercial model to examine whether refactoring correctness trends remain consistent across proprietary LLMs.

    \item Qwen2.5-Coder-32B-Instruct~\cite{qwen2025qwen25technicalreport} is included as a modern open-source, code-specialized model, enabling comparison with a strong coding-focused model that is more accessible for local or reproducible deployment than proprietary systems.
\end{itemize} 
The selection of four LLMs allows us to compare LLM-based refactoring across open-source, open-weight, and closed-source frontier models.

\subsection{Hardware Environment}
All experiments were conducted on a cluster of 8 NVIDIA A100 GPUs, each with 80 GB of memory, running on Ubuntu 22.04.4 LTS.

 \section{Results}
\label{sec:results}
In this section, we provide the motivation, approach, and results of our research questions.

\subsection{\textbf{RQ1: How often does LLM-based refactoring lead to incorrect code behavior?}}
\subsubsection{\textbf{Motivation}}
Prior studies have shown that LLM-generated code can introduce functional errors despite appearing plausible or improving code quality attributes~\cite{cordeiro2024empirical, xia2023keep, 10172854}. However, most existing work has focused on code generation or bug fixing, rather than systematically evaluating whether LLM-generated refactorings preserve behavior, which is a goal of code refactoring~\cite{fan2023large,chen2021evaluatinglargelanguagemodels}. A transformation that improves structure but changes external functionality cannot be considered a correct refactoring. Therefore, this research question investigates how often refactorings generated by LLMs lead to behavioral regressions, as measured by unit test outcomes.

\subsubsection{\textbf{Approach}}
To evaluate LLM-based refactoring, we generate refactorings using multiple LLMs under consistent prompting configurations, verify that the outputs constitute valid refactorings, and evaluate their functional correctness using unit test outcomes. We then apply statistical tests to analyze differences in pass rates across models, refactoring scopes, and prompting settings.

\noindent\textbf{Model Selection.} 
We evaluate StarCoder2~\cite{lozhkov2024starcoder2stackv2}, GPT-4o~\cite{openai2024gpt4}, Claude Sonnet~\cite{Anthropic2024Claude3.5Sonnet}, and Qwen2.5-Coder-32B-Instruct~\cite{qwen2025qwen25technicalreport} to cover LLMs with different levels of capability, accessibility, and specialization as mentioned in Section~\ref{sec:refactoring_generation}.

\noindent\textbf{Functional Correctness Evaluation.} 
To address this research question, we evaluate the functional correctness of code refactorings generated by each selected LLM on individual Java methods extracted from the target repositories, as described in Section~\ref{sec:test_method_extraction}. We report the overall pass rate (\ref{equ:pass rate}) as well as pass rates across method-level and class-level refactorings to understand how frequently LLM-generated changes maintain functional behavior. 

\noindent\textbf{Prompting Configuration.} 
To examine the effect of in-context examples, we run four prompting settings: 0-shot, 1-shot, 3-shot, and 5-shot. In the few-shot settings, examples are randomly sampled from the same project but are disjoint from the target method, and no retrieval or similarity-based selection is applied. Each shot count corresponds to the number of worked refactoring examples included in the prompt before the target method. We use the same prompt template across all evaluated models to ensure a consistent comparison. Decoding parameters, including temperature, top\_p, and maximum token limit, are held fixed within each model configuration to reduce variation caused by sampling and to isolate the impact of model choice and shot count.

To this end, refactorings are produced using the instruction prompt shown in Figure~\ref{fig:model_prompt}, which consists of a system-level instruction defining refactoring as a behavior-preserving code transformation~\cite{fowler1999refactoring}, followed by the target code snippet and a placeholder for the refactored output. Few-shot exemplars, when used, precede the target snippet in the same format.

\begin{figure}[t]
    \centering
    \begin{minipage}{0.80\linewidth}
        \centering
        {\footnotesize
        \fbox{
            \parbox{\textwidth}{
                \textbf{Instruction}\\
                You are a model specialized in refactoring Java code. 
                Refactoring improves the internal structure, readability, and 
                maintainability without changing external behavior. 
                Output a refactored version of the input code.\\[0.3cm]
                \textbf{Prompt}\\
                \textbf{\# unrefactored code snippet (java):}\\
                \{code\_segment\_before\_refactoring\}\\[0.3cm]
                \textbf{\# refactored version of the same code snippet:}
            }
        }}
    \end{minipage}
    \caption{Zero-shot instruction used to prompt each evaluated LLM for refactoring.}
    \label{fig:model_prompt}
\end{figure}

\noindent\textbf{Output Verification.} 
Before measuring test outcomes, we verify that each LLM actually performs refactoring rather than reproducing the original code. To do this, we use RefactoringMiner 2.0 (RMiner)~\cite{tsantalis2020refactoringminer}, a state-of-the-art refactoring detection tool that is widely used in empirical software engineering studies and supports the detection of method- and class-level refactorings. For each generated file, we compute a unified diff between the LLM-refactored version and the original version and run RMiner to detect structural refactorings. A generated output is considered a valid refactoring only if RMiner identifies at least one refactoring operation, such as \texttt{Rename Method}, in the diff between the original code and the LLM output. Outputs where RMiner detects no refactoring are excluded from the analysis to avoid counting identical or near-identical reproductions as successful refactorings.

\noindent\textbf{Statistical Evaluation of Refactoring Outcomes.} 
After filtering non-refactored cases, we compare Pass@1 rates across the evaluated LLMs, where Pass@1 denotes the percentage of generated refactorings that successfully compile and pass all associated unit tests on the first attempt. We first compare model-level pass-rate distributions to determine whether some LLMs preserve behavior more reliably than others. We then compare pass rates between method-level and class-level refactorings using the Mann-Whitney U-test~\cite{doi:https://doi.org/10.1002/9780470479216.corpsy0524}. This non-parametric test evaluates whether two independent samples differ significantly in their median success rates without assuming a normal distribution~\cite{doi:https://doi.org/10.1002/9780470479216.corpsy0524}.

To assess whether pass rates differ across the 0-shot, 1-shot, 3-shot, and 5-shot prompting settings, we apply the Kruskal-Wallis test~\cite{doi:https://doi.org/10.1002/9780470479216.corpsy0491}, a non-parametric alternative to one-way ANOVA, to determine whether at least one group differs significantly from the others. We report the test statistic, \(p\)-value, and effect size \((\hat{\epsilon}^2)\) to quantify the magnitude of the overall difference. When significant differences are observed across models or prompting settings, we apply a post-hoc multiple-comparison procedure, such as Dunn's test~\cite{repec:tsj:stataj:v:15:y:2015:i:1:p:292-300} with Holm correction, to identify which groups differ from one another.

\subsubsection{\textbf{Findings}}
\begin{figure}[t]
\centering
\includegraphics[width=0.70\linewidth]{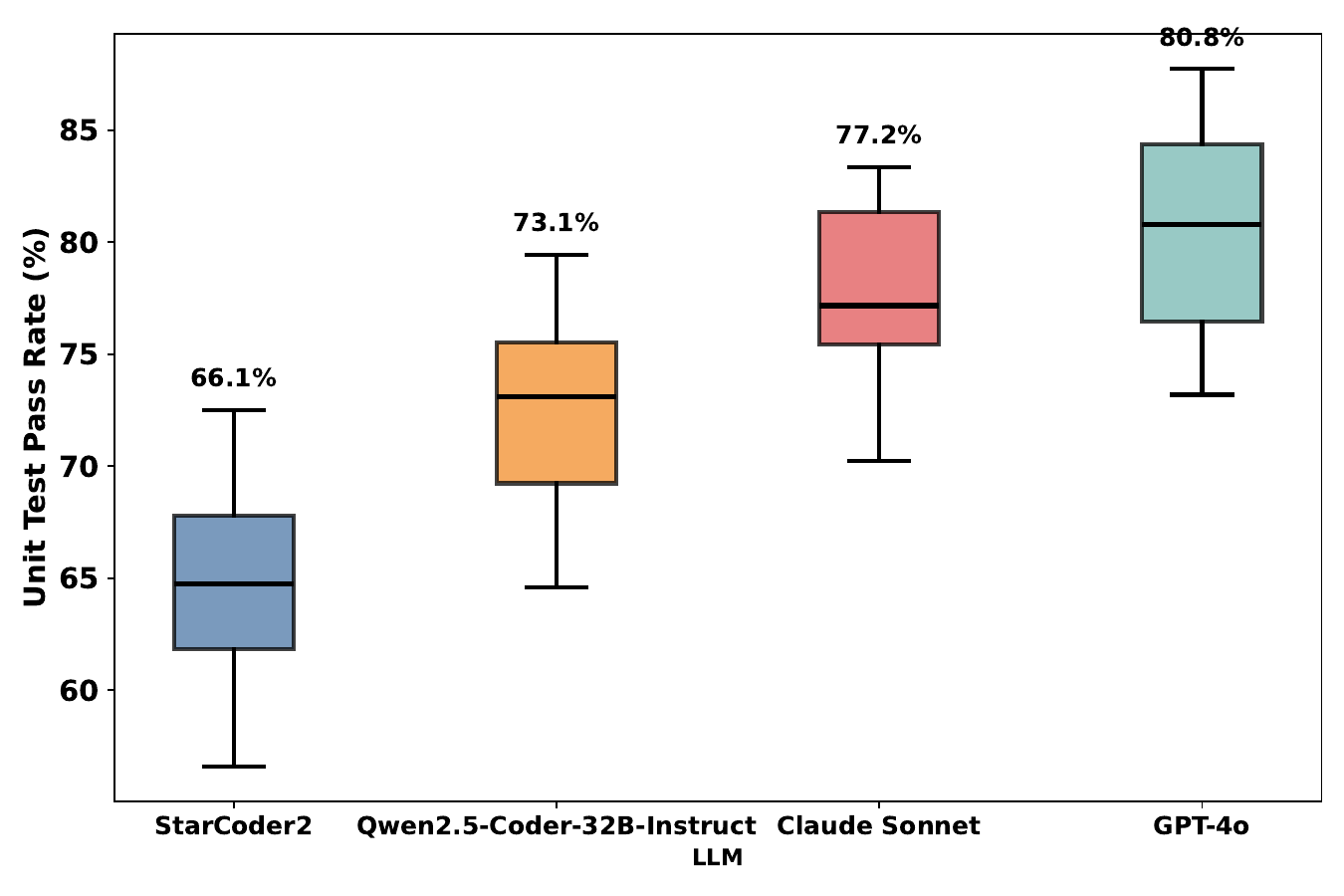}
\caption{Distribution of unit-test pass rates across evaluated LLMs.}
\label{fig:rq1_model_pass_rates}
\end{figure}

\begin{figure*}[t]
\centering
\includegraphics[width=0.90\linewidth]{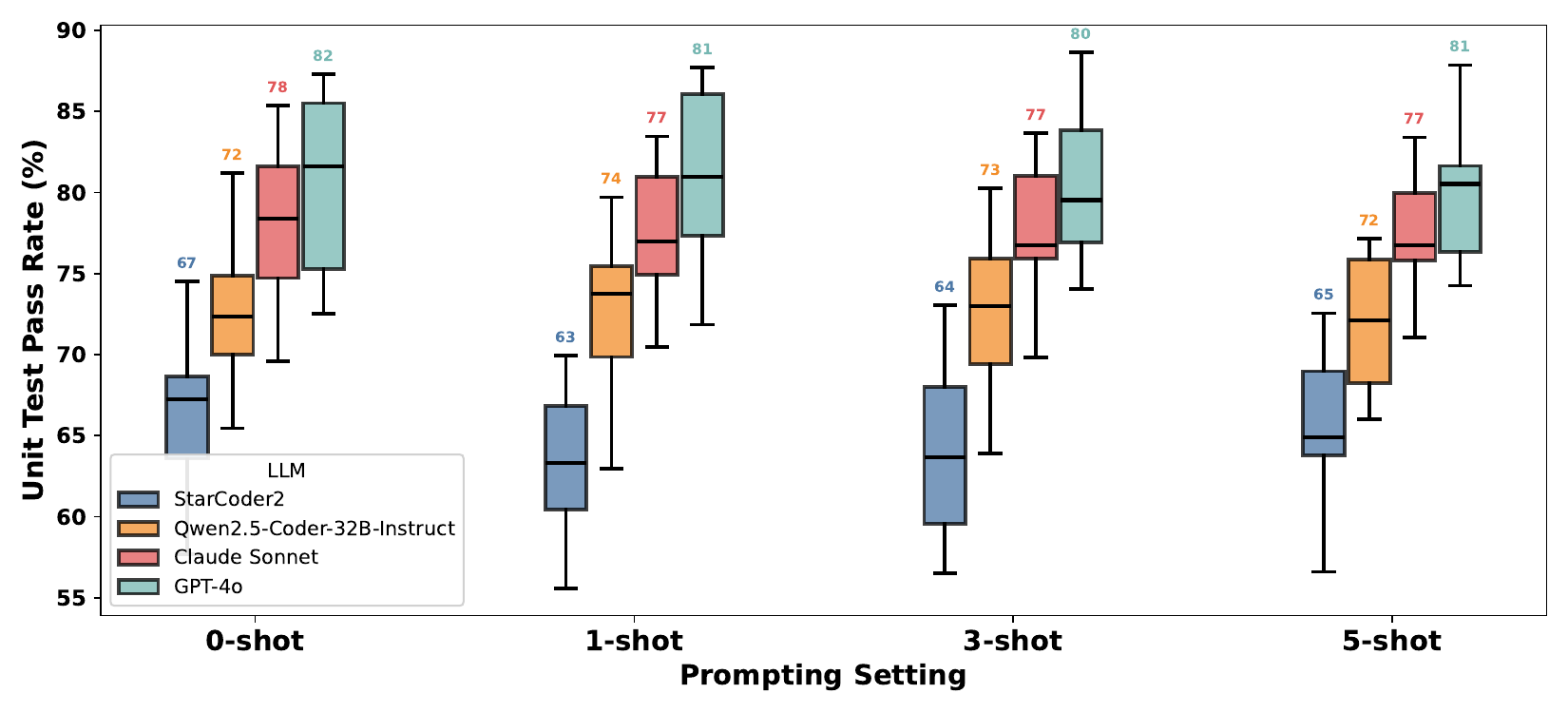}
\caption{Distribution of unit-test pass rates across prompting settings.}
\label{fig:pass_rate}
\end{figure*}

\begin{figure}
\centering
\includegraphics[width=0.70\linewidth]{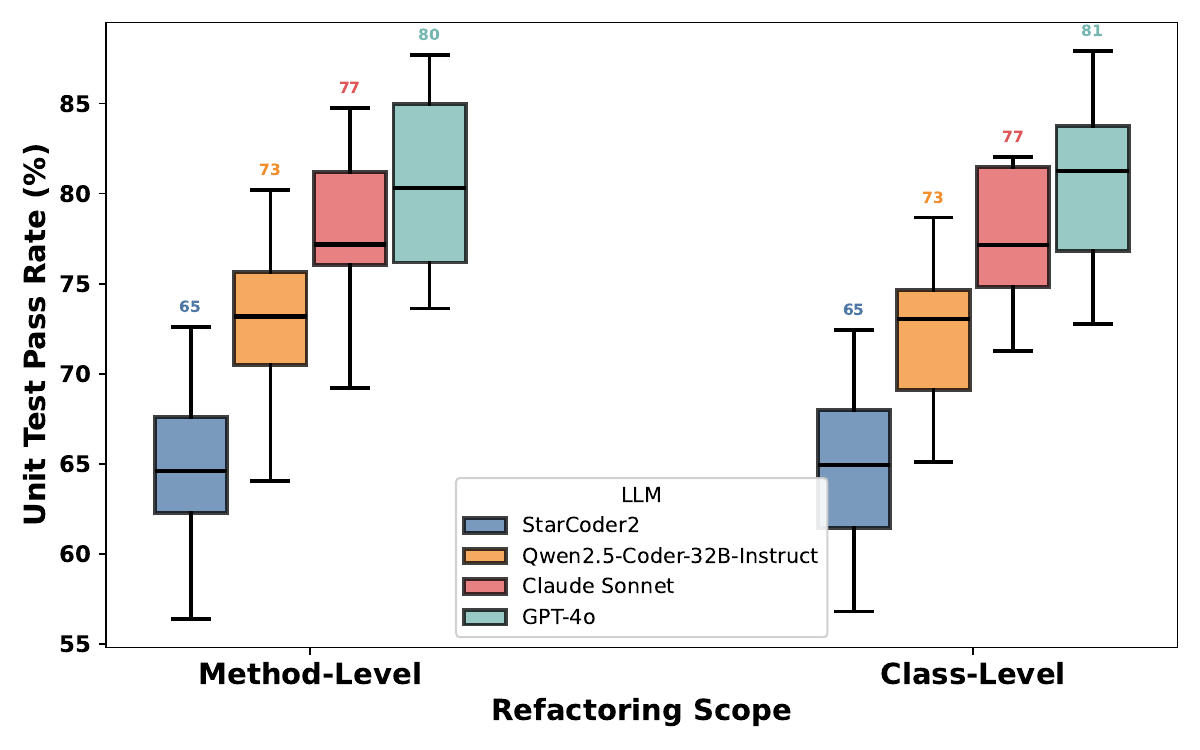}
\caption{Distribution of unit-test pass rates for method-level and class-level refactorings.}
\label{fig:method_class_pass}
\end{figure}

\textbf{LLM-generated refactorings frequently introduce behavioral regressions, even across models with larger parameter sizes and coding capabilities.}
Figure~\ref{fig:rq1_model_pass_rates} shows the distribution of project-level Pass@1 rates across the evaluated LLMs. The median Pass@1 rates are 66.1\% for StarCoder2, 80.8\% for GPT-4o, 77.2\% for Claude Sonnet, and 73.1\% for Qwen2.5-Coder-32B-Instruct. These results show a clear model effect: frontier closed-source models (GPT-4o, Claude Sonnet) generally achieve higher pass rates than open models, while StarCoder2 is consistently lowest. However, even the strongest model still exhibits failures 19.2\% of the time, indicating that behavior preservation remains a challenge for LLM-based refactoring. Although we control for this risk for StarCoder2 by excluding repositories included in its training data, the same level of verification is not possible for closed-source models because their training data is not fully disclosed. Therefore, the higher pass rates of GPT-4o and Claude Sonnet may be due to the exposure of the models to target or similar code, or a combination of both. 

The statistical comparison across models confirms significant differences. The Kruskal--Wallis test shows a significant difference in Pass@1 distributions across the evaluated LLMs (\(H = 25.18\), \(p = 1.4 \times 10^{-5}\), \(\hat{\epsilon}^{2} = 0.62\)). Dunn's post-hoc test with Holm correction shows that GPT-4o significantly outperforms Qwen2.5-Coder-32B-Instruct (\(p = 0.039\)) and StarCoder2 (\(p < 0.001\)), and Claude Sonnet significantly outperforms StarCoder2 (\(p < 0.001\)). GPT-4o and Claude Sonnet are not significantly different (\(p = 0.485\)), and the difference between Claude Sonnet and Qwen2.5-Coder-32B-Instruct is not significant (\(p = 0.143\)). These findings suggest that models with stronger coding and reasoning capabilities improve functional correctness, but do not eliminate behavioral regressions.

\textbf{Few-shot examples have a limited and inconsistent impact on functional correctness.}
Figure~\ref{fig:pass_rate} shows pass-rate distributions across 0-shot, 1-shot, 3-shot, and 5-shot prompting. The median project-level pass rates are 75.3\%, 73.6\%, 73.3\%, and 73.6\%, respectively. Overall, differences are small, and the grouped plot indicates no consistent monotonic gain from adding more examples across models. This observation is supported by the Kruskal--Wallis H-test, which reveals no statistically significant difference among the four prompting settings (\(H = 0.25\), \(p = 0.969\), \(\hat{\epsilon}^{2} \approx 0\)). These results suggest that correctness is driven more by underlying model capability than by simply increasing in-context examples.

\textbf{Method-level and class-level refactorings exhibit comparable functional correctness.}
Figure~\ref{fig:method_class_pass} compares Pass@1 rates between method-level and class-level refactorings. The median Pass@1 rate for method-level refactorings is 73.8\%, while class-level refactorings achieve 74.1\%. The Mann--Whitney U-test shows no statistically significant difference between the two groups (\(U = 50.0\), \(p = 1.00\)), indicating that refactoring scope does not substantially influence functional correctness.

This result suggests that failures are not explained solely by whether edits are confined to a method or span a class. Instead, regressions likely arise from deeper issues such as incomplete context understanding, inconsistent renaming, type mismatches, or unintended behavioral changes.

\medskip
\noindent
\fbox{%
  \parbox{0.97\linewidth}{%
    Our results show that LLM-based refactoring can preserve functional correctness, but behavioral regressions remain common across all evaluated models. Model choice has the strongest effect, with GPT-4o achieving the best overall functional correctness at an 80.8\% pass rate; however, no model fully eliminates compilation or unit-test failures. In contrast, few-shot prompting and refactoring scope have limited and less consistent effects on correctness, suggesting that downstream validation and repair remain necessary for reliable LLM-based refactoring.
  }%
}

\subsection{\textbf{RQ2: What are the most common reasons for failures in LLM-refactored code?}}
\subsubsection{\textbf{Motivation}}
In the first research question, we find that a median of only 66.1\% of StarCoder2-generated refactorings preserve functional correctness, and that neither few-shot prompting nor refactoring scope substantially improves unit-test pass rates. These results indicate that a significant portion of refactorings still introduce behavioral errors that are not mitigated by prompt-level adjustments alone. Therefore, understanding the root causes of unit test failures is critical to enhance LLM performance and testing methodologies. In this research question, we aim to analyze patterns in test failures and identify specific limitations in the LLM-refactored code.

\subsubsection{\textbf{Approach}}
To investigate common reasons for test failures in LLM-refactored code, we combine automated large-scale analysis with manual validation to identify recurring failure patterns and categorize them into a fixed taxonomy. For refactorings that fail unit tests, we analyze compilation errors and runtime test failures to identify common failure patterns. This analysis is performed on all failed refactorings to systematically categorize the root causes of failure. We then quantify the frequency of each failure category to understand the dominant failure modes of LLM-generated refactorings. These insights are further used to inform the design of our static-fix pipeline and guide the iterative repair process.

\noindent\textbf{Automated Failure Analysis.}
We employ an automated categorization step using GPT-4~\cite{openai2024gpt4}, as it has demonstrated strong performance on code understanding and natural language reasoning tasks, to analyze all unit test failures. This process consists of the following stages:

\begin{enumerate}
    \item LLM-based Summarization: For each failing case, GPT-4 produces a single-sentence natural language summary of the error log~\cite{panthaplackel-etal-2020-learning, tao2021evaluation}.
    
    \item Embedding and Clustering: We encode all GPT-generated summaries using the \textit{all-MiniLM-L6-v2} SentenceTransformer~\cite{wang2020minilmdeepselfattentiondistillation}. The resulting embeddings are clustered using KMeans~\cite{hartigan1979algorithm}.
    
    \item Cluster Optimization: To identify the most meaningful grouping of failures, we compute silhouette scores~\cite{rousseeuw1987silhouettes, noei2023empirical} across candidate cluster counts (2–12). The silhouette score measures how similar each data point is to others within its cluster compared to those in other clusters; higher scores indicate well-separated, coherent clusters. The best score (\textit{i.e.}, 0.34) occurs with eight clusters.
\end{enumerate}

\noindent\textbf{Manual Verification.} To evaluate the effectiveness of our automated failure analysis, we randomly select a statistical representative sample size of 350 failures, which corresponds to a 95\% confidence level and a 5\% margin of error~\cite{xu2024dataefficientevaluationlarge, shao2024balanceddatasamplinglanguage, noei2023empirical, yu2026empirical}. To this end, we select up to 35 unit test failures from each repository (for a total of 350) in our dataset to ensure balanced coverage across projects. As some repositories contain fewer than 35 failures, the final evaluation set consists of 300 unit test failures in total, while still preserving diversity across refactoring contexts and functional requirements. To validate the failure categories identified through automated analysis, we follow prior work~\cite{yu2026empirical, ahmed2025can} by employing a human evaluator and comparing the resulting categorizations with those produced by the LLM. Specifically, the first author independently analyzes each instance in the sampled set and determines its likely failure cause based on the refactored code, error logs, and test outcomes. Since our analysis is exploratory, neither the human evaluator nor the LLM assigns labels from a fixed, predefined set at the outset. Instead, both produce failure explanations that are subsequently normalized into a shared failure taxonomy for comparison.

When the manual evaluation label corresponds to the same failure category as our automated approach, the case is treated as an agreement; otherwise, it is treated as a disagreement. To quantify the level of agreement between the automated and manual analyses, we compute Cohen’s kappa~\cite{cohen1960coefficient}, which measures inter-rater agreement beyond chance. A high kappa score indicates that the automated clustering and labeling process is well aligned with the manually derived categories and therefore provides reliable support for identifying recurring test failure patterns at scale.

\subsubsection{\textbf{Findings}}
\label{sec:rq2_findings}

The manual analysis and alignment between the manually assigned labels and the GPT-4-assisted labels yielded a Cohen’s kappa of 0.78, indicating a high level of agreement~\cite{cohen1960coefficient}. This result suggests that the automatically identified clusters are largely consistent with the manually derived failure categories, supporting the reliability of our taxonomy for characterizing common unit test failure patterns in LLM-refactored code.

\begin{figure} [t]
    \centering
    \includegraphics[width=0.70\linewidth]{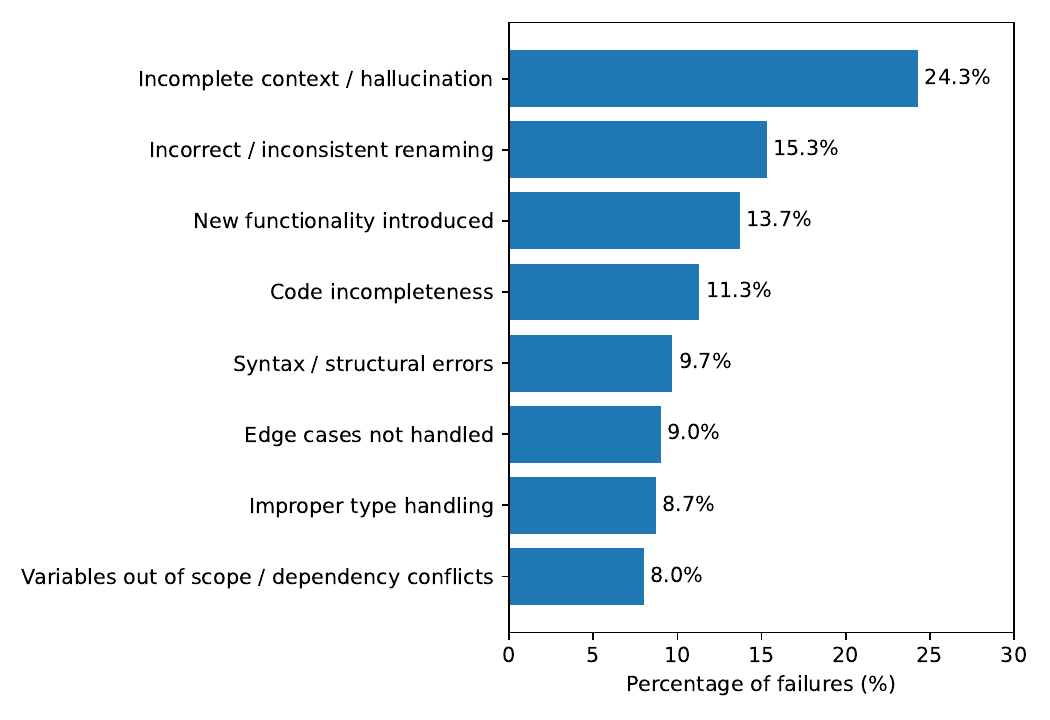}
    \caption{Breakdown of the reasons for unit test failures from manual analysis.}
    \label{fig:failure_reasons}
\end{figure}

\textbf{Many unit test failures stem from contextual misunderstandings and code-generation errors rather than semantic defects.} The results of our manual evaluation reveal insights into the common reasons for unit test failures in LLM-refactored code. The findings are summarized in Figure~\ref{fig:failure_reasons}, which illustrates the percent breakdown of the different failure reasons. Each failure reason is described in detail below.

\begin{itemize}
\item\textbf{Incomplete context or hallucination} is identified as the leading cause of failures, accounting for 24.3\%. This category highlights the LLM’s inability to fully understand the broader context of the codebase during refactoring. Failures in this category often involved inappropriate or unrelated changes, such as introducing unnecessary methods or class members that were inconsistent with the original functionality. Such hallucinations reflect limitations in the model’s ability to synthesize complex dependencies and maintain coherence within the refactored code. 

\item\textbf{Incorrect or inconsistent renaming} contributes to 15.3\% of test failures. These errors occurred when the LLM renamed variables, methods, or other identifiers inconsistently across the codebase. Inconsistent renaming resulted in broken references, logical mismatches, or runtime errors, particularly in cases where the renamed entities were used extensively within the code.

\item\textbf{Code incompleteness} is observed in 11.3\% of the analyzed failures. These issues are characterized by refactored code outputs that are syntactically incomplete, such as missing return statements, incomplete control flow structures, or undefined variables. Such failures indicate that the model sometimes terminates refactorings prematurely or lacks the necessary mechanisms to ensure completeness in its output. 

\item\textbf{Edge cases not handled} account for 9\%. These issues arise when the refactored code does not account for specific edge cases, leading to incorrect functionality under certain input conditions. This highlights the model’s inability to generalize effectively to less common scenarios, often due to a lack of training data covering diverse use cases.

\item\textbf{Syntax and structural errors} are responsible for 9.7\% of failures. These errors included basic violations such as unmatched parentheses, invalid syntax, or improper formatting.

\item\textbf{Adding new functionality or variables} accounts for 13.7\% of failures. This category involves the unintended introduction of elements that deviate from the original functionality, often leading to test failures due to unforeseen behavioral changes. These issues underscore the need for stricter constraints on the scope of LLM-generated refactorings to ensure alignment with the original code intent. 

\item\textbf{Variables outside scope or conflicts} account for 8\% of the total unit test failures analyzed. Failures in this category arise when variables are accessed outside their defined scope or when refactorings introduce conflicts with existing dependencies. These issues often result from the LLM’s lack of awareness of scoping rules and project-level dependencies during refactoring. 

\item\textbf{Improper type handling} contributes to 8.7\% (\textit{i.e.,} 26 out of 300 cases) of failures. These failures typically involve mismatched data types in variable assignments or method signatures, leading to compilation or runtime errors. These issues indicate that while LLMs are generally capable of inferring data types, they often struggle with maintaining type safety in complex contexts.
\end{itemize}

\medskip
\noindent
\fbox{%
  \parbox{0.97\linewidth}{%
    Our analysis shows that unit test failures in LLM-refactored code are dominated by contextual misunderstandings, hallucinated changes, and structural issues such as inconsistent renaming and incomplete code. These failure modes indicate that many incorrect refactorings arise from limitations in context comprehension and output completeness rather than from complex semantic errors, motivating targeted mitigation strategies such as static fixes and feedback-driven repair mechanisms.
  }%
}

\subsection{\textbf{RQ3: Can our agentic approach improve the correctness of LLM-refactored code?}}
\subsubsection{\textbf{Motivation}}
Our findings from the second research question reveal that many incorrect LLM-generated refactorings fail due to syntactic and shallow structural issues, such as missing imports, unbalanced brackets, improper type handling, and incomplete code fragments. These failure modes suggest that a substantial portion of unit test failures arises from easily identifiable and potentially correctable issues introduced during refactoring generation. However, static corrections alone are insufficient to resolve all failing refactorings, as some failures stem from deeper semantic inconsistencies, contextual misunderstandings, or incorrect transformations that require LLM reasoning. Therefore, in this research question, we investigate whether an agentic approach that integrates static fixes and an iterative repair loop can improve the functional correctness of LLM-generated refactorings. By applying rule-based corrections before each compilation and test cycle, and then invoking targeted LLM-based repair only when needed, our approach aims to recover incorrect refactorings while minimizing unnecessary edits and computational overhead.

\subsubsection{\textbf{Approach}}
To evaluate whether RefactorAssist improves the correctness of LLM-generated refactorings, we apply it to refactorings that fail compilation or unit testing after the initial generation stage. As described in Section~\ref{sec:method}, RefactorAssist first applies static fixes to address common syntactic and structural issues. Refactorings that still fail are then passed to the iterative agentic repair loop, where compiler diagnostics, unit-test failures, code diffs, retrieved project context, and diagnostic explanations are used to guide repair generation.

\textbf{Repair Generation Model.} We use StarCoder2~\cite{lozhkov2024starcoder2stackv2} as both the initial refactoring-generation model and the repair-generation model inside RefactorAssist. This setup helps us control for the effect of model substitution, ensuring that any improvement in correctness is attributed to RefactorAssist's static fixes, agentic repair, and repair guidance rather than to replacing the original model with a stronger code generation model.

\textbf{Error Diagnostic Explanation Models.} For the error diagnostic explanation component, we evaluate three LLMs: GPT-4o~\cite{openai2024gpt4}, Claude 3.5 Sonnet~\cite{Anthropic2024Claude3.5Sonnet}, and Llama-3 8B~\cite{grattafiori2024llama3herdmodels}. These models are used only to generate error explanations and repair hints; they do not directly edit the code. We select these models to compare diagnostic performance across different levels of capability, scale, and accessibility. GPT-4o and Claude 3.5 Sonnet provide strong reasoning ability and large context windows for handling long error traces and multi-file repair contexts, while Llama-3 8B serves as a lightweight, locally deployable open-source alternative. We report the results of all three models in the same evaluation.

To evaluate the performance of the repair pipeline, we measure the unit test pass criteria defined in the first research question. In addition, we record the iteration at which each successful repair occurs and track the cases that remain unresolved after the iteration limit, allowing us to assess both repair effectiveness and repair efficiency over time.

\subsubsection{\textbf{Findings}}

\begin{figure}[t]
  \centering
  \includegraphics[width=0.55\columnwidth]{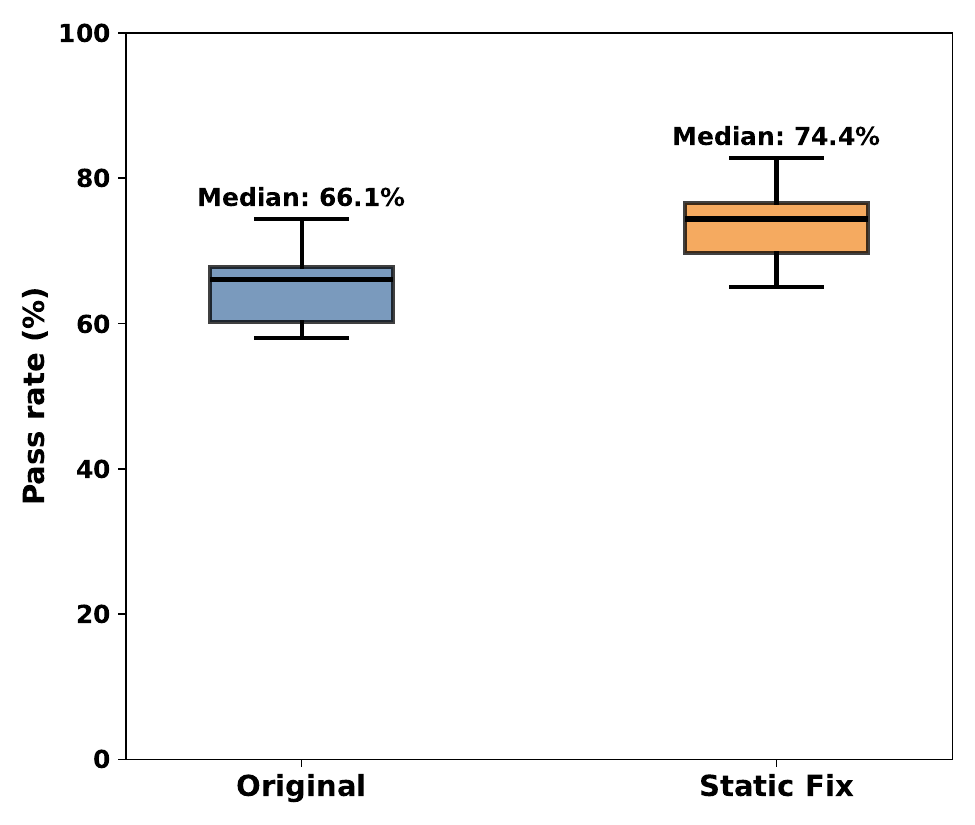}
  \caption{Project-level unit test pass rates for original and static-fixed refactorings. Boxes show the distribution across projects.}
  \label{fig:rq2_pass_rates_comparison}
\end{figure}

\textbf{Static fixes increase the median functional correctness of LLM-generated refactorings by 8.3 percentage points before entering the agentic refactoring repair stage of RefactorAssist.} Figure~\ref{fig:rq2_pass_rates_comparison} compares project-level unit test pass rates for original and static-fixed refactorings. Across projects, the median pass rate increases from 66.1\% in the Original setting to 74.4\% after applying static fixes, corresponding to a median improvement of 8.3 percentage points. After applying the static-fix step in RefactorAssist, 80.3\% of all refactorings pass the unit tests. The remaining 19.7\% of refactored methods and classes still fail and therefore proceed to the agentic refactoring repair stage of RefactorAssist.

\begin{figure}[t]
  \centering
  \includegraphics[width=0.70\columnwidth]{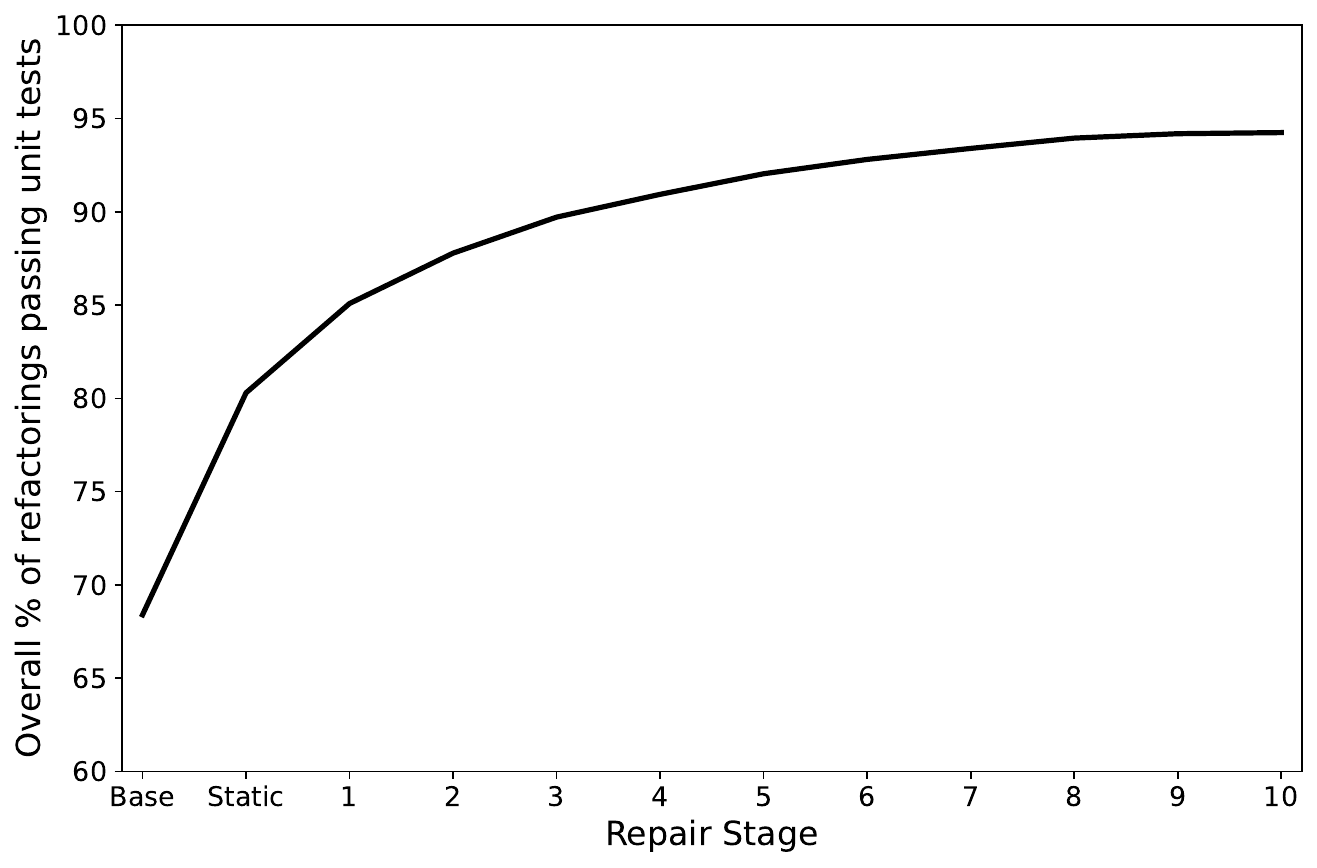}
  \caption{Overall unit test pass rate across the full refactoring repair pipeline. The first point shows the base pass rate of the original LLM-generated refactorings, the second shows performance after the static-fix stage, and the remaining points show the cumulative improvement after each iteration of the agentic refactoring repair stage.}
  \label{fig:rq3_feedback_loop_trendline}
\end{figure}

\textbf{The combined static-fix and iterative repair pipeline substantially improves the correctness of failing LLM-generated refactorings, reaching 93.4\% pass rate after 10 iterations.} Figure~\ref{fig:rq3_feedback_loop_trendline} shows the cumulative percentage of failures repaired by iteration \(k\) using the full repair pipeline, which combines static fixes, GPT-4o-based error diagnosis, retrieved project context, code diffs, error logs, and StarCoder2-based repair. The base pass rate of the original LLM-generated refactorings is 68.4\%. After the static-fix stage, this increases to 80.3\%. The remaining failures then enter the agentic refactoring repair stage, which raises the overall pass rate further over successive iterations, reaching 85.1\% after one iteration, 87.8\% after two iterations, 89.7\% after three iterations, and 93.4\% after ten iterations.

\textbf{Most successful repairs occur in the early iterations of the agentic refactoring repair stage.}
The largest gains occur within the first three repair attempts, after which the rate of improvement slows substantially. This pattern suggests that many failures can be resolved once the model is provided with focused diagnostic evidence and minimal repair guidance, while later iterations mainly address a smaller subset of harder cases. The diminishing gains after the third iteration also support the use of a bounded repair budget, as most practically recoverable failures are resolved early in the process.

\textbf{Remaining failures are likely associated with deeper semantic inconsistencies.}
Although RefactorAssist substantially improves correctness, a portion of failures (e.g., 6.6\%) remains unresolved after the iteration limit of 10. Consistent with the failure analysis in the second research question, these likely correspond to cases involving broader contextual misunderstandings, inconsistent renaming across scopes, hallucinated functionality, or other semantic issues that cannot be resolved through lightweight structural correction and localized iterative repair alone.

\medskip
\noindent
\fbox{%
  \parbox{0.97\linewidth}{%
    The combined static-fix and agentic refactoring repair stage substantially improves the functional correctness of LLM-generated refactorings. The overall unit test pass rate increases from 68.4\% for the original refactorings to 80.3\% after static intervention, and further rises through iterative repair to 93.4\%. Most successful repairs occur within the first three iterations, indicating that targeted diagnostic feedback is effective for recovering a large portion of incorrect refactorings.
  }%
}

\subsection{\textbf{RQ4: Which components of our approach contribute most to repair effectiveness?}}
\subsubsection{\textbf{Motivation}}
In the third research question we observe that the combined static-fix and agentic refactoring repair stage can substantially improve the correctness of failing LLM-generated refactorings. However, the repair pipeline consists of several interacting components, including code diffs, compiler and runtime error logs, explanation-LLM output, and retrieval-augmented project context. It remains unclear which of these components are most responsible for the observed repair gains, and whether stronger diagnosis LLMs consistently provide better repair guidance. Understanding the relative contribution of each component is important for both practical deployment and future agent design, as it helps identify which sources of feedback are essential and which provide only limited or situational benefit. Therefore, in this research question, we conduct an ablation study to determine how different diagnostic inputs and explanation LLMs affect repair effectiveness.

\subsubsection{\textbf{Approach}}
To understand which components of our repair pipeline drive the largest gains, we perform an ablation study over the four main inputs to the repair prompt:
\begin{enumerate}
    \item code diff,
    \item compiler/runtime error log,
    \item explanation-LLM output, and
    \item retrieval-augmented project context.
\end{enumerate}

For the explanation step, we evaluate three diagnosis LLMs: GPT-4o~\cite{openai2024gpt4}, Claude 3.5 Sonnet~\cite{Anthropic2024Claude3.5Sonnet}, and Llama-3 8B~\cite{grattafiori2024llama3herdmodels}. These models are used exclusively to produce structured failure explanations and fix hints; they do not directly edit the code. 

We evaluate all combinations of the four repair inputs and record, for each configuration, the percentage of failing refactorings repaired by iteration \(k\). We also record the number of iterations required for each successful repair and track the failures that remain unresolved after the iteration limit. To compare configurations statistically, we apply the Scott-Knott ESD~\cite{scott-knott, ouf2026empirical} test at the 0.05 level to the per-project fix-rate distributions~\cite{cohen1988statistical}. This test partitions the evaluated configurations into statistically distinct performance groups, allowing us to identify which combinations of diagnostic inputs yield significantly different levels of repair effectiveness while accounting for multiple comparisons.

\subsubsection{\textbf{Findings}}
\begin{table*}[t]
\centering
\caption{Ablation Study: Cumulative percent of initial errors fixed by iteration \(k\) for each feedback-loop configuration (10 iterations).}
\label{tab:rq3-ablation-errors-cum}
\resizebox{\textwidth}{!}{
\begin{tabular}{llrrrrrrrrrr}
\toprule
\multicolumn{2}{c}{} & \multicolumn{10}{c}{\textbf{Cumulative errors fixed by iteration \(k\) (\%)}} \\
\cmidrule(lr){3-12}
\textbf{Error Diagnostic Agent} & \textbf{Step} & \textbf{1} & \textbf{2} & \textbf{3} & \textbf{4} & \textbf{5} & \textbf{6} & \textbf{7} & \textbf{8} & \textbf{9} & \textbf{10} \\
\midrule
GPT--4o           & Diff + ErrLog + Expl             & 24.3 & 38.0 & 47.8 & 54.0 & 59.6 & 63.5 & 66.5 & 69.3 & 70.5 & \textbf{70.8} \\
Claude 3.5 Sonnet & Diff + ErrLog + Expl             & 22.9 & 37.3 & 46.0 & 52.9 & 57.0 & 60.5 & 63.1 & 65.3 & 66.5 & 66.7 \\
Llama 3 (8B)      & Diff + ErrLog + Expl             & 19.6 & 31.7 & 39.6 & 44.7 & 49.1 & 52.1 & 54.5 & 56.4 & 57.9 & 58.1 \\
GPT--4o           & Diff + ErrLog + Expl + RAG       & 21.7 & 34.9 & 43.8 & 50.6 & 55.8 & 59.9 & 62.8 & 65.1 & 67.0 & 67.3 \\
Claude 3.5 Sonnet & Diff + ErrLog + Expl + RAG       & 22.4 & 35.0 & 43.1 & 49.1 & 53.9 & 57.6 & 60.4 & 62.5 & 64.0 & 64.2 \\
Llama 3 (8B)      & Diff + ErrLog + Expl + RAG       & 18.7 & 29.9 & 36.8 & 42.2 & 45.8 & 48.5 & 50.7 & 52.5 & 53.6 & 53.7 \\
                  & Diff + ErrLog                    & 14.8 & 23.3 & 28.3 & 31.8 & 34.5 & 36.3 & 37.9 & 38.9 & 39.5 & 39.7 \\
                  & Diff + RAG                       & 13.1 & 20.3 & 25.8 & 29.0 & 31.8 & 33.9 & 35.3 & 36.3 & 36.8 & 36.9 \\
GPT--4o           & Expl only                        & 12.8 & 19.3 & 24.2 & 27.3 & 29.8 & 31.5 & 32.7 & 33.8 & 34.4 & 34.6 \\
                  & RAG only                         &  8.7 & 13.7 & 17.3 & 19.8 & 21.6 & 23.1 & 24.2 & 25.0 & 25.5 & 25.6 \\
                  & Diff only                        &  8.9 & 13.4 & 16.2 & 18.3 & 20.0 & 21.3 & 22.4 & 23.1 & 23.5 & 23.6 \\
\bottomrule
\end{tabular}}
\vspace{0.2cm}

\small \emph{Step legend:} Diff = code diff; ErrLog = compiler/runtime error log; Expl = agent-generated explanation and fix hint; RAG = retrieval-augmented context.

\small \emph{Notes:} Baseline pass rate before entering the feedback loop is 80.3\%, so the initial error pool is 19.7\% of methods; percentages are reported relative to this pool. Rows labeled with \textbf{Expl} use an LLM as an explanation agent (GPT-4o, Claude 3.5 Sonnet, or Llama 3 (8B)). Rows without \textbf{Expl} are signal-only baselines that do not invoke an explanation agent and are included to isolate the contribution of each feedback signal. Iteration 10 contributes $\approx$0\% additional fixes due to diminishing returns.
\end{table*}

\begin{figure}[t]
  \centering
  \includegraphics[width=0.80\columnwidth]{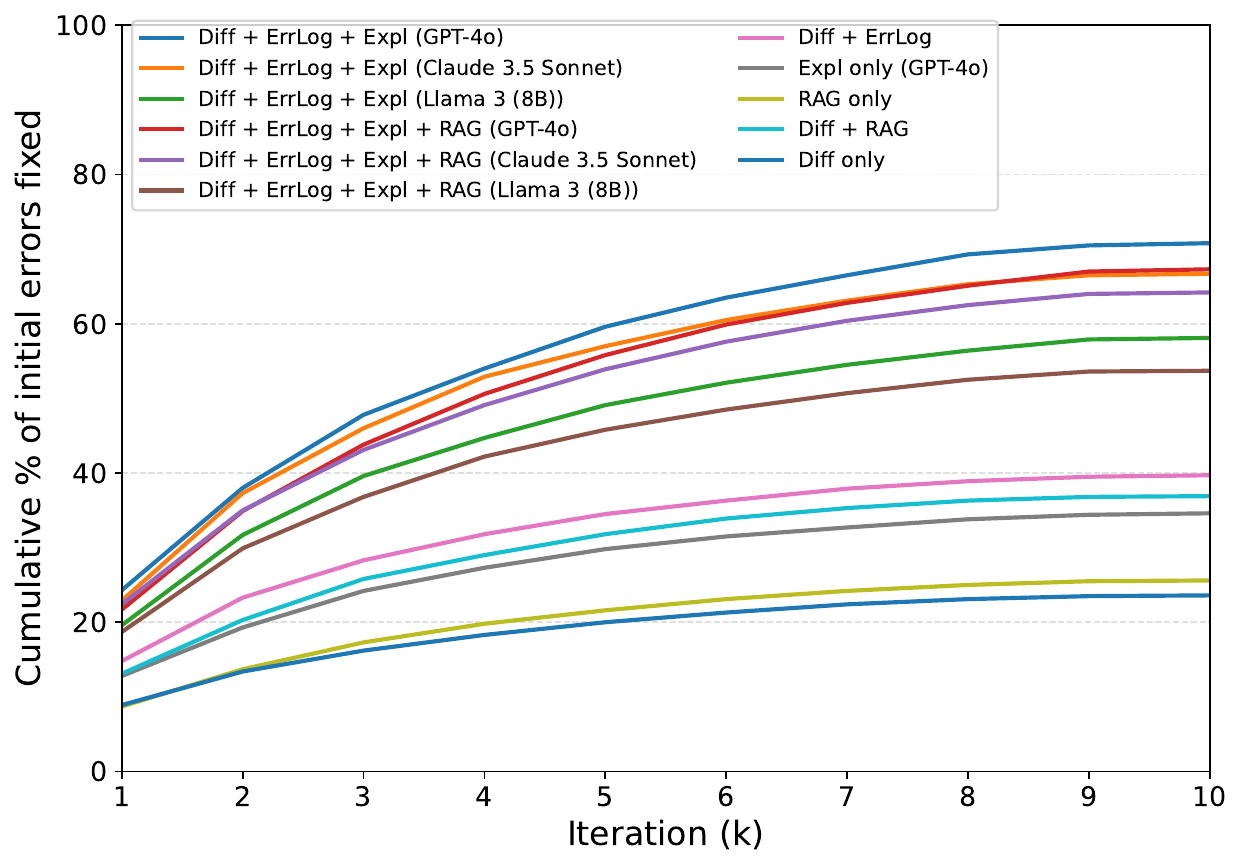}
  \caption{Cumulative percentage of failures remaining after the static-fix stage repaired by iteration \(k\) across ablation configurations. These failures correspond to the 19.7\% of refactorings that still fail after static intervention.}
  \label{fig:ablation_trendlines}
\end{figure}

\textbf{Explanation-aware configurations achieve the strongest repair performance, with the best configuration repairing 70.8\% of failures remaining after static intervention for a total refactoring success rate of 94.2\%.}
The cumulative results in Table~\ref{tab:rq3-ablation-errors-cum} and Figure~\ref{fig:ablation_trendlines} show that configurations combining code diffs, error logs, and explanation-LLM output consistently achieve the highest repair rates. The values in Figure~\ref{fig:ablation_trendlines} are computed over the failures that remain after the static-fix stage, which correspond to 19.7\% of all refactorings. Among the remaining failures, the strongest configuration, which uses the code diff, error log, explanation output, and GPT-4o as the error diagnostic LLM, repairs 24.3\% in the first iteration and 38.0\% by the second iteration. By the third iteration, it fixes 47.8\% of all recoverable failures, and after ten iterations reaches a cumulative recovery rate of 70.8\%. These results indicate that explicit failure explanation is a critical ingredient in effective iterative repair.

\textbf{The quality of the error diagnostic LLM strongly influences repair effectiveness.} Claude 3.5 Sonnet and Llama 3 (8B) exhibit the same overall trend but achieve lower final recovery rates. Specifically, Claude 3.5 Sonnet reaches 66.7\% after ten iterations, while Llama 3 (8B) recovers 58.1\% of errors under the same configuration. The differences indicate that stronger explanation LLMs produce more accurate root-cause diagnoses and repair hints, which directly improve downstream patch generation. 

\textbf{Retrieval-augmented generation accelerates early fixes but is not sufficient for the final recovery.} Adding retrieval (``+RAG'') yields modest gains in the first one to two iterations (e.g., 21.7\% vs.\ 24.3\% at iteration one without RAG) but results in lower final coverage (67.3\% vs.\ 70.8\%). This suggests that retrieval primarily assists in quickly resolving familiar or surface-level issues, while later iterations benefit more from precise failure explanations than from additional context retrieval. Adding retrieval-augmented context does not yield statistically significant gains in final recovery. For GPT-4o, the Diff + ErrLog + Expl configuration (mean 70.8\%) is statistically indistinguishable from its RAG-augmented variant (67.3\%), confirming that retrieval primarily accelerates early fixes rather than increasing overall repair coverage. A similar pattern holds for Claude 3.5 Sonnet, where adding RAG shifts the configuration from Rank 1 to Rank 2.

\begin{table}[t]
\scriptsize
\centering
\caption{Scott--Knott ESD grouping of feedback-loop configurations based on fixable percentage (\%).  
Lower ranks indicate significantly better performance (Rank 1 is the best rank).}
\label{tab:rq3-sk-summary}
\begin{tabular}{lrrr}
\toprule
\textbf{Configuration} & \textbf{Mean} & \textbf{Median} & \textbf{SK Rank} \\
\midrule
Diff + ErrLog + Expl (GPT-4o)                  & 70.8 & 66.15 & \textbf{1} \\
Diff + ErrLog + Expl + RAG (GPT-4o)            & 67.3 & 71.20 & \textbf{1} \\
Diff + ErrLog + Expl (Claude 3.5 Sonnet)       & 66.7 & 67.30 & \textbf{1} \\
\midrule
Diff + ErrLog + Expl + RAG (Claude 3.5 Sonnet) & 64.2 & 66.60 & \textbf{2} \\
\midrule
Diff + ErrLog + Expl (Llama 3 (8B))            & 58.1 & 56.40 & \textbf{3} \\
Diff + ErrLog + Expl + RAG (Llama 3 (8B))      & 53.7 & 58.95 & \textbf{3} \\
\midrule
Diff + ErrLog                                  & 39.7 & 38.40 & \textbf{4} \\
\midrule
Diff + RAG                                     & 36.9 & 38.30 & \textbf{5} \\
Expl only (GPT-4o)                             & 34.6 & 33.90 & \textbf{5} \\
\midrule
RAG only                                       & 25.6 & 26.30 & \textbf{6} \\
\midrule
Diff only                                      & 23.6 & 23.65 & \textbf{7} \\
\bottomrule
\end{tabular}
\end{table}

\textbf{Explanation-aware configurations significantly outperform all other agent designs.}
Table~\ref{tab:rq3-sk-summary} reports the Scott–Knott ESD grouping of agentic refactoring repair stage configurations based on the percentage of fixable errors. The top-ranked group (Rank 1) consists of configurations that combine code diff, error logs, and explanation-LLM output: GPT-4o (with and without RAG) and Claude 3.5 Sonnet without RAG. These configurations achieve the highest repair rates, with mean fixable percentages ranging from 66.7\% to 70.8\%, and are statistically superior to all other variants based on the Scott-Knott ESD test. \textbf{Weaker explanation LLMs lead to substantially lower repair rates.} Configurations using Llama 3 (8B) consistently occupy Rank 3, with mean fixable percentages between 53.7\% and 58.1\%, indicating a clear drop in effectiveness compared to GPT-4o and Claude 3.5 Sonnet. This gap highlights the role of high-quality failure explanations in guiding successful repairs.

\textbf{Partial or single-input configurations perform poorly.} Ablations that omit the explanation LLM, such as Diff + ErrLog (39.7\%), Diff + RAG (36.9\%), or Expl-only (34.6\%), fall into lower ranks (Ranks 4-5). Minimal configurations using only RAG or only the diff achieve the lowest repair rates (25.6\% and 23.6\%, respectively), demonstrating that no single component is sufficient for effective automated repair. Finally, the sharp drop in marginal improvements after the first few iterations, together with the dominance of explanation-aware configurations, suggests that most fixable errors are shallow and diagnostic-driven, whereas remaining failures likely require deeper semantic reasoning or human intervention.

\medskip
\noindent
\fbox{%
  \parbox{0.97\linewidth}{%
    RefactorAssist can substantially improve the functional correctness of failing LLM-refactored code, with most recoverable errors fixed within the first three iterations. RefactorAssist configurations that combine code diffs, error logs, and the error diagnostic LLM consistently achieve the highest repair rates, with the highest being \textbf{70.8\%} with GPT-4o as the error diagnostic LLM. The effectiveness of the repair process is strongly influenced by the quality of the explanation LLM, indicating that accurate failure diagnosis is critical for successful automated repair. 
  }%
}
\section{Threats to Validity}
\label{sec:threats_to_validity}
This section outlines potential threats to the validity of our study and describes the measures taken to mitigate them.

\subsection{Threats to Internal Validity}
Although we filter overlapping repositories that are excluded from the Stack-v2 dataset and not included in StarCoder2's training data, it is possible that the model learned general patterns or styles from similar codebases. To mitigate this risk, we select 10 projects with distinct characteristics, ensuring minimal similarity with common training examples.
Another threat is the inherent challenge of hallucinations in LLM-generated refactorings. These hallucinations may introduce syntactically valid but semantically incorrect changes. To address this issue, all refactored code is validated using unit tests, and failure cases are analyzed to identify and categorize recurring patterns. The mapping of test cases to methods depends on heuristic and static analysis techniques, which may overlook certain relationships, particularly in cases involving dynamically generated code or complex dependency injection frameworks. To reduce this risk, we compile and run the corresponding unit test for each method to ensure its validity before adding it to our dataset.
Lastly, the selection of commits and repositories could introduce bias. Although we randomize the selection process within the curated dataset, the exclusion of certain repositories based on filtering criteria may inadvertently favor simpler codebases or more well-documented projects. To mitigate this, we ensure a representative distribution of projects across various domains, sizes, and complexities.

\subsection{Threats to External Validity}
The generalizability of our findings is limited to the scope of the study, which focuses on open-source Java projects with sufficient commit history and test coverage. These findings may not directly extend to other programming languages or proprietary software systems. Furthermore, our evaluation is based on a specific version of StarCoder2, and results may vary with updates to the model architecture or future LLMs. Additionally, the dataset primarily includes repositories with robust test coverage, which may not reflect real-world scenarios where test cases are incomplete or absent. This could lead to an overestimation of the model’s performance in environments with limited testing infrastructure. 

Our evaluation is also limited by the scope of the refactoring targets. Since Methods2Test~\cite{Tufano_2022} provides method-level mappings between production code and unit tests, we restrict the evaluated refactoring targets to single-method and single-class changes so that behavioral outcomes can be attributed more directly to the intended transformation. This restriction helps reduce unrelated cross-file edits, build configuration changes, or modifications outside the mapped method-test relationship. However, this design may limit generalizability to multi-file refactorings, where correctness often depends on interactions among several classes, interfaces, and imports. To mitigate this limitation, RefactorAssist is not restricted to the target method or class when diagnosing failures; when compiler errors, test failures, or code diffs indicate cross-file dependencies, it can use repository context during iterative refinement.

To mitigate these threats, we evaluate the model on a diverse set of real-world projects from multiple domains, use repositories with established development histories and unit tests to ensure reliable behavioral assessment, and report our experimental setup in sufficient detail to support reproducibility and future replication across other models, languages, and software settings.

\subsection{Threats to Construct Validity}
We use unit test pass rates as the primary metric for evaluating functional correctness. While this metric provides an objective measure, it may not fully capture other aspects of code quality, such as readability, maintainability, or adherence to design principles. Subtle issues related to long-term code maintenance are not directly assessed in this study.

Our statistical analyses, including significance testing and effect size reporting, rely on the sample size and distribution of test outcomes. Small sample sizes in specific failure categories or project domains could reduce the statistical power of our conclusions. To address this, we ensure adequate sample sizes and report p-values to provide clarity.

Finally, the automated extraction of test-method mappings introduces potential errors. Any inaccuracies in the mapping process could affect the evaluation of refactorings. Because Methods2Test~\cite{Tufano_2022} links methods to associated unit tests, our evaluation assumes that the mapped tests are suitable for checking whether the refactored code behaves correctly. Inaccurate or incomplete mappings could therefore affect whether a passing test outcome fully reflects functional correctness. To mitigate this, we validate the mappings in our dataset by compiling and running the associated tests for each target before evaluation to exclude cases where the original method-test pair is not executable.

\subsection{Implications}
Our findings provide several implications for practitioners, researchers, and tool developers working with LLM-based code refactoring.

\textbf{Implications for Practitioners.}
Our results show that while LLMs can produce code refactorings, they frequently introduce functional regressions. This suggests that LLM-generated refactorings should not be directly integrated into production systems without validation. Practitioners should incorporate automated testing and validation pipelines when adopting LLM-assisted refactoring tools. Our findings also highlight the role of static fixes, which can eliminate common syntactic and structural issues before testing and thereby improve the correctness of generated refactorings at low cost without the invocation of LLMs. In addition, the strong gains achieved by the iterative repair mechanism suggest that agent-based validation and repair can substantially improve correctness with limited manual intervention. In a broader multi-agent software engineering ecosystem, our approach can therefore serve as a dedicated refactoring agent that proposes quality-improving transformations while coordinating with testing agents to preserve functional behavior.

\textbf{Implications for Researchers.}
This study highlights the gap between structural improvements and functional correctness in LLM-generated refactorings. Future research should focus on improving the reliability of LLMs in preserving program semantics, particularly in multi-file and dependency-aware settings. Our results also suggest that combining static fixes with iterative agentic repair is a promising direction for improving robustness, as these stages address different classes of failures and together substantially improve functional correctness. The observed failure patterns further provide a foundation for developing more targeted benchmarks, taxonomies, and evaluation frameworks for refactoring tasks. More broadly, our findings suggest that refactoring should be studied not only as a one-step code generation problem, but as a multi-stage workflow involving generation, validation, diagnosis, and repair. This opens opportunities for future work on agent-based refactoring systems, better retrieval mechanisms for project-specific context, and training or fine-tuning strategies tailored to semantics-preserving code transformation rather than general-purpose code generation.

\textbf{Implications for Tool Developers.}
The effectiveness of static fixes and iterative repair pipelines suggests that LLM-based refactoring tools should be designed as hybrid systems rather than standalone generators. Instead of relying on a single model invocation, tool builders should integrate compiler feedback, test execution, error diagnosis, and repair into the refactoring workflow to improve output reliability. Our findings also suggest the value of modular architectures in which different models or components can be assigned specialized roles, such as initial refactoring, failure diagnosis, and targeted repair. This design enables flexibility across deployment settings while making it easier to incorporate retrieval mechanisms, project-specific context, and validation feedback into the overall refactoring pipeline.

\section{Related Work}
\label{sec:related_work}
This section reviews key areas of related work, focusing on automated refactoring, testing LLM-generated code, and approaches to repairing failing refactorings.

\textbf{Static Code Refactoring.} Automated code refactoring tools aim to improve software quality by restructuring code without altering its functionality~\cite{fowler2018refactoring}. Early refactoring tools, such as JRefactory~\cite{1138} and Eclipse’s Refactoring Engine~\cite{10.5555/1197540}, are rule-based, relying on static analysis to apply predefined transformations. While these tools effectively handle straightforward tasks such as renaming variables or extracting methods, they often require manual intervention to address more complex, context-dependent refactoring scenarios.

In contrast to these rule-based approaches, our work evaluates LLM-generated refactorings at the method-level and class-level, and focuses on producing refactorings as well as whether their functional correctness can be preserved and recovered through automated repair.

\textbf{Automated Code Refactoring with LLMs.} The advent of LLMs, such as Codex~\cite{chen2021evaluatinglargelanguagemodels}, GPT-4~\cite{openai2024gpt4}, and StarCoder2~\cite{lozhkov2024starcoder2stackv2}, has enabled more flexible and intelligent refactoring solutions. These models generate human-like code and structural changes, offering new possibilities for automated code improvement, including tasks such as Extract Method refactoring guided by LLM suggestions~\cite{10.1145/3663529.3663803}. However, studies highlight that while LLMs excel in reducing code smells and simplifying logic~\cite{cordeiro2024empirical, cordeiro2025llm}, they often fail to preserve semantic correctness, frequently producing hallucinated or incorrect transformations that lead to unit test failures~\cite{10479398}.

Unlike prior studies that primarily assess whether LLMs can generate refactorings or reduce code smells, our work systematically examines their behavioral correctness through unit testing and further investigates how incorrect refactorings can be recovered using static fixes and an iterative repair pipeline.

\textbf{Testing LLM-Generated Code.} Unit testing plays a critical role in validating the functional correctness of LLM-generated refactorings. Tools such as EvoSuite~\cite{10.1145/2025113.2025179} generate test cases that aim to achieve high structural coverage, but they may not capture the requirements of real-world scenarios. The disparity between automatically generated and manually developed tests is a key challenge in assessing the reliability of LLM-generated refactorings~\cite{Brownlee_2020}.

Research has also explored integrating LLMs into Test-Driven Development (TDD) workflows. For instance, \textit{LLM4TDD} leverages LLMs to generate code from test cases, demonstrating improved code quality when high-quality tests are provided~\cite{piya2023llm4tddbestpracticestest}. The approach introduced by LLM4TDD can be integrated into an iterative feedback loop to generate refactorings based on unit tests. Mutation testing, another evaluation strategy, has been used to identify flaws in refactored code by injecting artificial defects and assessing whether the test suite detects them~\cite{PAPADAKIS2019275}.

Building on this line of work, our study uses unit testing not only as an evaluation mechanism for LLM-generated refactorings, but also as an integral feedback signal that drives iterative diagnosis and repair of incorrect transformations.

\textbf{LLM-Based Refactoring Repair.} While LLMs can propose meaningful refactoring changes, they frequently introduce errors that require manual correction. This limitation has spurred research into automated refactoring repair mechanisms, which aim to identify and fix issues in LLM-generated refactorings. Automated Program Repair (APR) techniques, such as those based on evolutionary algorithms~\cite{7886945, le2011genprog} and neural repair models~\cite{gupta2017deepfix}, have been adapted to validate and refine LLM-generated code. These methods typically focus on applying minimal changes to pass failing tests, aligning closely with the goals of this study.

Shirafuji et al.~\cite{10479398} explore leveraging few-shot prompting to improve the quality of LLM-generated refactorings. While effective, their approach relies heavily on prompt design and does not dynamically adapt to real-world test results. Similarly, Liu et al.~\cite{liu2024empiricalstudypotentialllms} propose using few-shot prompts to guide LLMs, significantly improving refactoring success rates but highlighting the need for iterative refinement to address persistent errors. 

Our work combines failure analysis, static correction, and an agentic refactoring repair stage to automatically diagnose and recover incorrect LLM-generated refactorings based on observed compilation and test outcomes.
\section{Conclusion}
\label{sec:conclusion}
This study examines the common problems in generating automated code refactorings using large language models and explores designing a refactoring repair agent for improving their functional correctness. Our investigation provides both empirical evidence of the current limitations in LLM-driven refactoring and enhancement strategies.
We find that StarCoder2 achieves a median project-level unit test pass rate of only 66.1\%, indicating that many generated refactorings fail to preserve intended functionality. Analysis of method versus class-level refactoring changes shows no statistically significant difference in pass rates, suggesting that refactoring scope alone does not meaningfully affect functional correctness. Furthermore, we show that the common failure patterns include incomplete context understanding, inconsistent renaming, and missing edge-case handling. To address these challenges, we design a two-stage repair strategy. We first apply static fixes that target syntactic and structural issues, increasing the median unit test pass rate from 66.1\% to 74.4\%. Second, we introduce RefactorAssist, which is an agentic approach that integrates unit test failure logs, code diffs, and LLM-generated explanations into the repair prompt. RefactorAssist uses the error diagnostic component and incorporates both the code diff and error log, which repairs up to 70.8\% of the initially failing refactorings after static treatment and produces a total cumulative pass rate of 94.2\%. These results advance the understanding of LLM-based refactoring by quantifying where current models succeed and fail, offering concrete targets for training data and algorithmic improvements. By identifying common refactoring failure patterns and demonstrating an effective agentic repair approach, this work lays the foundation for more robust refactoring assistants that can be reliably deployed in real-world software engineering pipelines.

Future work will explore integrating static analysis and semantic reasoning tools into the agentic refactoring repair stage, fine-tuning LLMs on refactoring-specific datasets, and scaling this framework to more diverse programming languages and project types.

\bibliographystyle{ACM-Reference-Format}
\bibliography{references}

\appendix

\end{document}